\documentclass[aps,prl,twocolumn,groupedaddress]{revtex4}
\usepackage{graphicx}
\begin{document}

\title{Zero Cosmological Constant from Normalized General
Relativity}

\author{Aharon Davidson}
\email[Email: ]{davidson@bgu.ac.il,rubinsh@bgu.ac.il}
\author{Shimon Rubin}
\affiliation{Physics Department, Ben-Gurion University,
Beer-Sheva 84105, Israel}
\date{July 21, 2009}

\begin{abstract}
	Normalizing the Einstein-Hilbert action by the volume
	functional makes the theory invariant under constant
	shifts in the Lagrangian.
	The associated field equations then resemble unimodular
	gravity whose otherwise arbitrary cosmological constant
	is now determined as a Machian universal average.
	We prove that an empty space-time is necessarily Ricci
	tensor flat, and demonstrate the vanishing of the
	cosmological constant within the scalar field paradigm.
	The cosmological analysis, carried out at the
	mini-superspace level, reveals a vanishing cosmological
	constant for a Universe which cannot be closed as long as
	gravity is attractive.
	Finally, we give an example of a normalized theory of
	gravity which does give rise to a non-zero cosmological
	constant. 
\end{abstract}

\maketitle

The combined physics of gravity and matter is conventionally
described by the theory of General Relativity (GR).
The latter stems from the Einstein-Hilbert action 
\begin{equation}
	I_{GR}=\int 
	\left(-{\cal R}-2\Lambda_{0}+{\cal L}_{m} \right)
	\sqrt{-g}~d^{4}x ~,
	\label{EH}
\end{equation}
which gives rise to the gravitational field equations
\begin{equation}
	{\cal R}_{\mu\nu}-\frac{1}{2}g_{\mu\nu}{\cal R}=
	\Lambda_{0} g_{\mu\nu} -\frac{1}{2}T_{\mu\nu}~.
	\label{EinsteinWithCosmo}
\end{equation}
Our unit convention is $16\pi G_{N}=1$ (and also $\hbar=c=1$).
The Ricci scalar $\cal R$ and the matter Lagrangian ${\cal L}_{m}$
are generically accompanied by a cosmological constant
$\Lambda_{0}$ term\cite{LambdaRev} which represents an
arbitrary constant curvature source.
Zel'dovich\cite{Zel'dovich} was the first to argue that the
cosmological constant $\Lambda_{0}$ consists in fact of
two pieces
\begin{equation}
	\Lambda_{0} =\Lambda _{bare}+
	\frac{1}{2} \rho _{vac} ~.
	\label{Zel'dovich}
\end{equation}
The bare cosmological constant $\Lambda _{bare}$, put by hand
into the Lagrangian, gets quantum mechanically shifted by the
vacuum expectation value of the energy density associated with
the matter fields.
Furthermore, even if one classically enforces $\Lambda _{bare}=0$,
the cosmological constant will re-appear at the semi-classical
level as the sum over all zero point energies of all normal modes. 

Assuming that the standard electro/nuclear gauge interactions
hold all the way to the Planck scale, then $10^{19}GeV$
becomes the natural ultraviolet energy cutoff for the quantum
effects entering $T_{\mu\nu}$.
Unfortunately, such a cutoff would give rise to a huge vacuum
energy density of $10^{76}~GeV^{4}$, roughly $123$ orders of
magnitude larger than the currently observed value of
$10^{-47}~GeV^{4}$.
Following conventional wisdom, the bare cosmological constant
$\Lambda_{bare}$ would then have to be fine-tuned to stunning $123$
decimal places, thereby constituting the so called cosmological constant
problem, one of the worst fine-tuning problems to ever appear in
theoretical physics.
Theories which invoke parameter adjustments of incredible accuracy
are usually referred to as "unnatural".
A first (perhaps necessary) step to render such a theory "natural" would
be, for example, to uncover an underlying \emph{symmetry principle}
capable of enforcing the tiny parameter of the theory, the cosmological
constant in our case, to be exactly zero.
This, however, will still leave the door open for the secondary puzzle,
namely the understanding why there is after all "something rather than
nothing".

Adopting the ${\cal L}\rightarrow{\cal L}+const$ symmetry principle
as the primary tool to make $\Lambda_{0} $ trivial, we hereby consider
the following so-called \emph{normalized} Einstein-Hilbert action
\begin{equation}
	I=\frac{I_{GR}}{\epsilon~I_{V}}=
	\frac{\int (-{\cal R}+{\cal L}_{m})\sqrt{-g}~d^{4}x}
	{\epsilon\int \sqrt{-g}~d^{4}x}~,
	\label{Action}
\end{equation}
where the standard Einstein-Hilbert functional $I_{GR}$ has
been divided by the volume functional $I_{V}$.
The above action was first introduced by Tseytlin \cite{Tseytlin}
in a keen attempt to solve the cosmological constant problem.
The idea was originally motivated as the effective low energy
limit of some duality symmetric closed string theory, but it is
quite obvious that it has life of its own, and should be regarded
as independent of any string theoretic underpinning.
In this paper, the action eq.(\ref{Action}) is regarded fundamental.
Note that $\Lambda_{0}$ has been absorbed, as designed,
with $I+2\Lambda_{0} /\epsilon$ serving now as a newly
defined $I$.
Given the fact that a constant shift in the numerator functional
cannot affect the local extrema, one may get the false impression
that the action eq.(\ref{Action}) is ignorant of the cosmological
constant.
However, we will soon explain that this is \emph{not} necessarily
the case.
We refer to the corresponding theory as normalized general
relativity (NGR).
The constant factor $\epsilon$, having units of $(length)^{-4}$,
has been introduced on purely dimensional grounds.
The two integrals better share a common domain of integration
over the space-time manifold.

On the list of other prominent attempts to resolve the cosmological
constant puzzle one can find:

\noindent
(i) \emph{Supersymmetry} - 
In theories where supersymmetry is unbroken, the net contribution
to $\left\langle\rho\right\rangle$ amounts to zero.
The fact is, however, that supersymmetry, if exists, must be broken,
with the associated breaking scale being at least at the $TeV$ level,
corresponding to a vacuum energy $\leq 10^{12}~ GeV^4$.
While this is a 64 orders of magnitude improvement in comparison
with non-supersymmetric theories, it is still 59 orders of magnitude
short.
In certain superstring models\cite{string}, the cosmological
constant might vanish even though supersymmetry gets broken.

\noindent
(ii) \emph{Alternative measures} -
In the Einstein-Hilbert action eq.(\ref{EH}), $\Lambda_{0} $ is the
constant coefficient of the standard measure $\sqrt{-g}$.
Being ready to deviate from GR, one may trade the
scalar density $\sqrt{-g}$ for a total derivative, namely
$\displaystyle{\sqrt{-g} \rightarrow
A^{\mu}_{~;\mu}\sqrt{-g}=(\sqrt{-g}A^{\mu})_{,\mu}}$.
The resulting theory stays invariant under the transformation
${\cal L}\rightarrow {\cal L}+const$, thereby turning $\Lambda_{0} $
into a locally irrelevant quantity.
A more sophisticated attempt in this category is provided by the
non-Riemannian total derivative measure density
$\frac{1}{4!}\epsilon^{\mu\nu\lambda\sigma}
\partial_{\mu}\phi_1\partial_{\nu}\phi_2\partial_{\lambda}
\phi_3\partial_{\sigma}\phi_4$ which carries degrees of freedom
$\phi_{i}$ independent of that of the metric tensor and the matter
fields.
A partial realization of this novel idea is achieved within the
framework so-called two-measure theory\cite{Guendelman}.

\noindent
(iii) \emph{Unimodular gravity} -
One way to ease the severeness of the cosmological constant
problem is to make $\Lambda_{0} $ a constant of integration.
This is the case in unimodular gravity\cite{unimodular} which
is characterized by the constraint $\sqrt{-g}=1$.
In this covariant (with a preferred volume element) theory,
since the action has to be stationary only with respect to
variations that keep $\sqrt{-g}$ fixed, i.e. for which
$g^{\mu\nu}\delta g_{\mu\nu}=0$, one obtains in place of the
full Einstein equations only their traceless $\Lambda_{0} $-free
part
\begin{equation}
	{\cal R}^{\mu \nu }-\frac{1}{4}g^{\mu \nu }{\cal R}=
	-\frac{1}{2}\left( T^{\mu \nu }-\frac{1}{4}
	g^{\mu \nu }T\right) . 
	\label{unimodular}
\end{equation}
These equations, first introduced by Einstein\cite{Einstein1} when
discussing the possible role played by gravitational fields in the
structure of elementary particles, give the false impression that
the trace $\cal R$ has been left out.
However, under the assumption that the energy-momentum
tensor is covariantly conserved (which is not guaranteed in
non-diffeomorphism invariant theories),
the Bianchi identities imply
\begin{equation}
	-{\cal R}_{,\mu }+\frac{1}{2}T_{,\mu}=0
	~~\Longrightarrow~
	-{\cal R}+\frac{1}{2}T=4\Lambda ~.
	\label{unimodularTrace}
\end{equation}
Remarkably, a newborn \emph{constant of integration} $\Lambda$
has replaced the original put-by-hand $\Lambda_{0}$, which has
disappeared from the equations, as the physical
cosmological constant.
Unfortunately, although the cosmological constant is now
determined by initial conditions, one still lacks a mechanism
to control its value.

Before diving into the associated equations of motion, one cannot
fail to notice that the above action is by definition non-local, at
least in the Gel'fand-Fomin\cite{Gel'fand} sense.
Counter intuitively, however, we will show that the associated
equations of motion are de facto local, such that the sole effect of
the apparent non-locality is expressed in this case by globally
fixing the value of a newly emerging $\Lambda$.
This may suggest the existence of an equivalent local theory
which is still at large.

Note that non-local functionals of this kind have been considered
in the literature, two examples of which are in order:

\noindent
(i) \emph{Degravitation} -
A phenomenological approach to the cosmological constant
problem, based on generally covariant non-local and acausal
modifications of GR, has been recently proposed\cite{degravitaion}.
The postulated equation of motion, lacking the support of an
underlying variation principle, involves a carefully designed
so-called 'filter function'.
In the far infrared, this filter function is expected to extract the zero
mode part of the Einstein tensor, which is proportional to $g_{\mu\nu}$,
and give rise to
\begin{equation}
	{\cal R}_{\mu \nu}-\frac{1}{2}g_{\mu \nu}{\cal R}=
	\frac{\eta^{2}}{4~}g_{\mu \nu }\left\langle {\cal R} \right\rangle
	-\frac{1}{2}T_{\mu \nu}  ~.
\end{equation}
The latter effective field equation involves, in some similarity with
our case, the space-time averaged Ricci curvature 
\begin{equation}
	\left\langle {\cal R} \right\rangle=
	\frac{\int {\cal R} \sqrt{-g}~d^{4}x}{\int \sqrt{-g}~d^{4}x} ~,
	\label{average}
\end{equation}
and also some large mass ratio $\eta$.

\noindent
(ii) \emph{Yamabe problem} - Invoking the non-local functional
\begin{equation}
	\frac{\int {\cal R} \sqrt{-g}~d^{n}x~~}
	{\left(\int \sqrt{-g}~d^{n}x \right)^{\frac{n-2}{n}}} ~, 
	\label{Yamabe}
\end{equation}
Yamabe\cite{Yamabe} has proven that any compact Riemannian
manifold of dimension $n\geq 3$ can be conformally mapped
into a constant scalar curvature manifold.
The power $\frac{n-2}{n}$ so chosen to guarantee the invariance
of eq.(\ref{Yamabe}) under global rescaling of the metric
$g_{\mu\nu}\rightarrow k g_{\mu\nu}$.
Interestingly, a signature reversal symmetry
$g_{\mu\nu}\rightarrow -g_{\mu\nu}$, a discrete version of
the latter, has been recently invoked\cite{signature} to tackle
the cosmological constant puzzle.

Given the action eq.(\ref{Action}), one first verifies that all classical
matter field equations and geodesic trajectories remain intact.
It is only at the semi-classical level, i.e. quantum matter fields in a
non-dynamical gravity background, that the $(\epsilon I_{V})^{-1}$
factor, which multiplies ${\cal L}$, is suppose to enter the game.
A tenable choice for $\epsilon$, soon to be discussed, can be made
at this level.

The gravitational field equations, on the other hand, derived by
varying the action eq.(\ref{Action}), are deceptively local
\begin{equation}
	{\cal R}_{\mu\nu}-\frac{1}{2}g_{\mu\nu}{\cal R}=
	\Lambda g_{\mu\nu} -\frac{1}{2}T_{\mu\nu} ~,
	\label{NGReq}
\end{equation}
where the constant $\Lambda\equiv \epsilon I/2$ is
still to be calculated.
The crucial point is that in NGR, unlike in unimodular
GR, $\Lambda$ does not stay arbitrary.
The reason is simple.
The solution $g_{\mu\nu}(x;\Lambda)$ of
eq.(\ref{NGReq}) can be re-used to actually calculate the value
$I(\Lambda)$ of the action along the classical path.
This way, $\Lambda$ must be a solution of the
functional equation
\begin{equation}
	I(\Lambda)=
	\frac{2\Lambda}{\epsilon} ~,
	\label{Ieq}
\end{equation}
to be regarded as a self-consistency condition.
Note that had we not absorbed $\Lambda_{0} $ in the first
place, then $\Lambda\equiv \epsilon I/2+\Lambda_{0} $,
leading to $I=2(\Lambda- \Lambda_{0} )/\epsilon$ which is
eq.(\ref{Ieq}) in disguise.
We find it remarkable that the entire non-locality of the theory
goes into fixing one single constant, namely $\Lambda$.

Together, eqs.(\ref{NGReq},\ref{Ieq}) are non-local
and acausal since they contain a space-time average of
Einstein-Hilbert action.
Yet, once $\Lambda$ gets fixed, eq.(\ref{NGReq})
becomes practically local.
Such a situation where the equations of motion keep a local
character, but have a 'minimal touch' of non-locality, can be
considered a realization of the Mach principle.
The latter refers to the vague hypothesis that \emph{''Mass
there governs inertia here''}, or as interpreted by Hawking
and Ellis\cite{HawkingEllis}, \emph{''Local physical laws are
determined by the large-scale structure of the Universe''}.
With this in mind, we recall the sensitivity of the action
eq.(\ref{Action}) to the domain of integration.
Following the Mach philosophy, the integration should be
carried out over the entire space-time manifold.

Naturally, this understanding poses a potential problem
for space-time manifolds of infinite volume.
One way out is to 'regularize' $\epsilon$, such that
$\epsilon V_{4}$ be finite, with the 4-volume $V_{4}$ denoting
the value of $I_{V}$ associated with a particular solution.
In particular, if QFT in a flat background is to be fully
recovered, i.e. governed by traditional
$\int {\cal L}\sqrt{-g_{flat}}~d^{4}x$ action, a tenable
choice for $\epsilon$ is $\epsilon V_{4}^{flat} \rightarrow 1$.

The strength of NGR can be demonstrated already at the level
of empty space-time.
First recall that any arbitrary $\Lambda_{0} $ we could have started
from gets eaten up by shifting $I$ by $2\Lambda_{0} /\epsilon$,
so the only question left is whether some newly emerging
$\Lambda$ makes its appearance or not.
Starting from a vanishing energy-momentum tensor
$T_{\mu\nu}=0$, the trace of eq.(\ref{NGReq}) implies
that the Ricci scalar must be a constant, 
${\cal R}=-4\Lambda$ to be specific.
One can then easily show that associated with some (still
arbitrary) $\Lambda$, is the value
$I(\Lambda)= 4\Lambda/\epsilon$
of the action.
Confronting the latter with the functional eq.(\ref{Ieq})
finally establishes one of our main results, namely
\begin{equation}
	\frac{4\Lambda}{\epsilon}=
	\frac{2\Lambda}{\epsilon}
	\quad\Longrightarrow\quad
	\Lambda=0 ~.
	\label{EmptyFlat}
\end{equation}
Within the framework of NGR,
and consistent with Einstein\cite{Pais} philosophy, \emph{an
empty space-time is necessarily Ricci tensor flat}, namely
$T_{\mu\nu}= 0 ~\Rightarrow~ \Lambda= 0$.
Bear in mind, however, that $T_{\mu\nu}\neq 0$
does not necessarily imply $\Lambda\neq 0$.

It is important to emphasize that deviating from general
relativity can drastically change eq.(\ref{EmptyFlat}), and
give rise to a non-trivial $\Lambda$, even in
the absence of matter fields.
Two examples are in order:

\noindent (i) The addition of a total derivative to the
Lagrangian, such as the Gauss-Bonnet term (in 4-dim)
\begin{equation}
	{\cal L}_{GB}=b\left(
	{\cal R}^{2}-4{\cal R}^{\mu\nu}{\cal R}_{\mu\nu}
	+{\cal R}^{\mu\nu\lambda\sigma}
	{\cal R}_{\mu\nu\lambda\sigma}\right) ~,
\end{equation}
while preserving the local equations of motion, will now
modify eq.(\ref{EmptyFlat}) into
\begin{equation}
	4\Lambda
	+\frac{8}{3} b\Lambda^{2}=
	2\Lambda~.
\end{equation}
$\Lambda=0$ is still a solution, but the other
one $\Lambda=-3/4b$, although misbehaving
for small $b$, makes us wonder whether the non-zero
value of the physical cosmological constant is of a
topological origin.

\noindent (ii) Certain so-called $f(R)$ theories
of gravity will give rise to $\Lambda\neq 0$.
An example is provided by the normalized action
\begin{equation}
	I=\frac{-1}{\epsilon\int \sqrt{-g}~d^{4}x}
	\displaystyle{\int 
	\left({\cal R}+\frac{\alpha}{{\cal R}}\right)
	\sqrt{-g}~d^{4}x}~.
	\label{1/R}
\end{equation}
Tracing the associated gravitational field equations,
one arrives at
\begin{equation}
	{\cal R}+\frac{3\alpha}{{\cal R}}
	+\nabla^{\mu}\nabla_{\mu}\frac{3\alpha}
	{{\cal R}^{2}}+4\Lambda=0 ~,
\end{equation}
with $\Lambda=\epsilon I/2$.
The constant curvature ${\cal R}$ solution, for which
$\nabla_{\mu}{\cal R}=0$, is then substituted back into
eq.(\ref{1/R}), leading us to
\begin{equation}
	\Lambda=-{\cal R}=\pm\sqrt{\alpha}~,
\end{equation}
admitting the proper $\alpha \rightarrow 0$ behavior.
This raises the intriguing possibility that a non-vanishing
cosmological constant actually signals a deviation from
normalized general relativity.
Self consistency requires $\alpha>0$.
This is to be contrasted with the underlying (non-normalized)
theory\cite{Carroll}, for which ${\cal R}=\pm\sqrt{-3\alpha}$,
and thus necessitates $\alpha<0$ (argued to be a source of
instability\cite{instability}, which we lack here).

To study the effect of matter in NGR, it seems pedagogical to
first derive the extension of Schwarzschild geometry
surrounding a point-like particle of mass $M$.
However, following the Geroch-Traschen
theorem\cite{GerochTraschen}, GR (and hence NGR) is
not capable of consistently dealing with co-dimension
$n \geq 2$ gravitating sources.
With this in mind, to probe the effect of matter we
(i) Calculate the space-time average
$\left\langle {\cal R} \right\rangle$ of the Ricci scalar
directly from the metric tensor in vacuum, and
(ii) Neglect the gravitational self-force\cite{selfF}
effects, and approximate the source contribution by
$-M\int dt$.

The spherically symmetric metric, static by virtue of Birkhoff
theorem, is of the generic Schwarzschild (anti) de-Sitter type
\begin{equation}
	ds^{2}=-F(r)dt^{2}+F^{-1}(r)dr^{2}+
	r^{2}d\Omega^{2}~,
	\label{metric}
\end{equation}
with $\displaystyle{F(r)=1-\frac{M}{8\pi r}-
\frac{1}{3}\Lambda r^{2}}$.
The corresponding ${\cal R}(\Lambda)$ is known
to be a constant everywhere, except for a Dirac delta function
contribution at the origin.
To be more specific, it is the Laplacian piece
$r^{-2}\partial_{r}\left(r^{2}\partial_{r}F(r)\right)$, residing
in the Ricci scalar ${\cal R}$, which gives rise (like in the weak
field limit) to the source $-M\delta(r)/8\pi r^2$, such that
\begin{equation}
	\int {-\cal R}\sqrt{-g}~d^{4}x=
	 4\Lambda V_{4}+\frac{1}{2}M\int dt ~.
	 \label{Mdt1}
\end{equation}
Altogether, collecting the various pieces which constitute
$I(\Lambda)$, one can verify that the
functional eq.(\ref{Ieq}) now reads
\begin{equation}
	\frac{4\Lambda}{\epsilon}-
	\frac{M\int dt}{2\epsilon V_{4}}=
	\frac{2\Lambda}{\epsilon} ~.
	\label{IMdt}
\end{equation}
Noticing that $V_{4}=V_{3} \int dt$ holds for an arbitrary
$\Lambda$, where in the 'flat' notation
$V_{3}=\frac{4}{3}\pi R^{3}$, we finally arrive at
\begin{equation}
	\Lambda=\frac{M}{4 V_{3}} ~.
	\label{MoverV}
\end{equation}
The fact that $M>0$ is crucial, forcefully implying
$\Lambda\geq 0$, thereby eliminating
the option of an anti de-Sitter background.
Moreover, the combination of a finite $M$ and an
\emph{infinite} $V_{3}$, which is the case here, clearly
dictates $\Lambda\rightarrow 0$, thus
singling out the Schwarzschild solution.
Intriguingly, exactly the same formula, namely
$\Lambda_{E}=4\pi G\rho$ in conventional units ($\rho$
denoting matter density), albeit for a \emph{finite} spatial
volume $V_{3}$, characterizes the static so-called Einstein
Universe.
This should be regarded a coincidence, and in particular, it
does not necessarily imply that a finite $\Lambda$
is correlated with the Einstein Universe.

A natural step now would be to consider the normalized
general relativity extension of FRW cosmology.
The trouble is that the standard FRW energy-momentum 
tensor is usually introduced at the level of the equations,
but is not always derivable directly from an underlying action
principle.
This difficulty can be effectively bypassed at the
mini-superspace level.
With the $3$-space integrated out, and the energy density
$\rho(a)$ playing the role of the potential part, one may start
from the tenable action
\begin{equation}
	I=\frac{2}{\epsilon\int a^3 dt}
	\int \left( 3\frac{\ddot{a}a+
	\dot{a}^2+k}{a^2}-\rho(a) \right)a^3 dt~.
	\label{mini}
\end{equation}
But now, unlike in the Hawking-Hartle analysis, the total
derivative $\frac{d}{dt}(6a^{2}\dot{a})$ cannot be eliminated
from the nominator.
Substituting the corresponding field equation
$\dot{a}^{2}+k=\frac{1}{3}(\rho+\Lambda)$
into eq.(\ref{mini}), and invoking the energy-momentum
conservation law $a\rho^{\prime}+3(\rho+p)=0$,
the functional consistency condition becomes
\begin{equation}
	\Lambda=\frac{1}{2}
	\left\langle\rho+3p\right\rangle 
	\quad \Longleftrightarrow \quad
	\langle \frac{\ddot{a}}{a} \rangle =0~. 
\end{equation}
The generic solution is again
$\Lambda \rightarrow +0$ (for a detailed
analysis see ref.\cite{nFRW}).
It primarily reflects the fact that $a\rightarrow \infty$ as
$t\rightarrow\infty$.
Furthermore, the solution associated with the attractive
gravity case $\rho+3p>0$ comes with a bonus, telling
us that the Universe cannot be spatially closed, i.e.
$k\leq 0$.
The only non-generic solutions, with
$\Lambda\neq 0$, are associated with $k>0$
Einstein-like Universes, such that
$sign(\Lambda)=sign(\rho+3p)$.

The simplest field theoretical example one can think
of involves a real scalar field $\phi(x)$, for which
\begin{eqnarray}
	&&{\cal L}_{\phi}=
	-\frac{1}{2}g^{\mu\nu}\phi_{,\mu}\phi_{,\nu}-
	V(\phi) ~,  \\
	&&T_{\mu\nu}=\phi_{,\mu}\phi_{,\nu}+
	g_{\mu\nu}{\cal L}_{\phi} ~.~~
\end{eqnarray}
Tracing and then space-time averaging the gravitational field
equations, as demonstrated by eq.(\ref{average}), gives rise to
\begin{equation}
	-\left\langle {\cal R} \right\rangle =
	4\Lambda+\frac{1}{2}
	\left\langle g^{\mu\nu}
	\phi_{,\mu}\phi_{,\nu}\right\rangle+
	2\left\langle V\right\rangle ~.
\end{equation}
One can now calculate the value of the action along the classical
solution, and find
\begin{equation}
	I(\Lambda)=
	\frac{4\Lambda}{\epsilon}+
	\frac{\left\langle V\right\rangle}{\epsilon} ~.
\end{equation}
In turn, the functional eq.(\ref{Ieq}) implies
$\Lambda+\frac{1}{2}\left\langle V\right\rangle=0$,
and the theory becomes equivalent to a general relativistic
minimally coupled scalar field theory governed by the effective
scalar potential
\begin{equation}
	V_{eff}(\phi)=V(\phi)-\left\langle V\right\rangle ~,
	\label{Veff}
\end{equation}
which is manifestly invariant under $V\rightarrow V+const$.

Of particular interest is the case where the potential is bounded
from below, that is $V(\phi)\geq V_{min}$.
The configuration with the lowest energy density is then
associated with $V(\phi(t))=V_{min}=\left\langle V\right\rangle$,
and is thus characterized, \emph{irrespective of the value of}
$V_{min}$, by $\Lambda=0$.
The latter result holds even in cases where $V_{min}$ is only
asymptotically approached at $t\rightarrow +\infty$.
Such an example is provided by an expanding FRW Universe
where the positive Hubble constant $H(t)=\dot{a}(t)/a(t)$ induces
a friction term in the scalar field equation, leading eventually to
\begin{equation}
	\left\langle V\right\rangle=
	\frac{\int V(\phi(t)) a(t)^{3} dt}{\int a(t)^{3} dt} 
	\rightarrow V_{min}  ~.
\end{equation}
In particular, this quintessence\cite{quintessence} category
does not exclude a short inflationary episode which only adds
a  finite contribution to the numerator.

Notice that had we traded eq.(\ref{Veff}) for the milder equation
$V_{eff} =V-V_{min}$, our main conclusion so far would have
remained unchanged.
But eq.(\ref{Veff}) is by far stronger!
Recalling that the cosmological constant (being a true
constant, in contrast with dark energy) is characterized by
being the space-time average of itself, and appreciating
the fact that in a scalar field theory, the cosmological
constant solely resides in the scalar potential, we are driven
to the conclusion that
\begin{equation}
	\Lambda=
	\frac{1}{2} \left\langle V_{eff}\right\rangle
	\equiv 0 ~.
	\label{Lambda0}
\end{equation}
We further argue that when adding the variety of standard
model fields into ${\cal L}_{m}$, eq.(\ref{Veff}) establishes
a simple yet powerful mechanism for canceling out their
vacuum's zero-point energy contribution to the cosmological
constant.
In fact, there is no need to calculate the zero-point energy
$\rho_{vac}$, or even correctly identify its physical cutoff.
The only vital piece of information is that it is a conserved
constant energy density.
This way, with $V_{eff}(\phi)$ being inert to
$V(\phi) \rightarrow V(\phi)+\rho_{vac}$, eq.(\ref{Lambda0})
prevails.

At any rate, a cosmological observer equipped with GR,
but being unaware of NGR, would presumably interpret
the slowly varying dark energy
\begin{equation}
	\Lambda_{eff}(t)=\frac{1}{2} V_{eff}(\phi(t)) \geq 0
	\label{Lambdaeff}
\end{equation}
as todays 'cosmological constant', and consistent with recent
observations, would find its tiny value to be positive definite.
Unfortunately, its exact value is beyond the scope of the
present approach (see \cite{CCvalue} for attempts to
yield the correct order of magnitude for the observed
'cosmological constant').

The overall message carried by
eqs.(\ref{EmptyFlat},\ref{MoverV},\ref{Lambda0},\ref{Lambdaeff}),
which characterizes NGR, is consistent
with Einstein's philosophy, and is expected to extend
beyond the special cases discussed.
To summarize our main points, they are:
(i) The cosmological constant is a concrete non-local
realization of Mach principle,
(ii) An empty space-time is necessarily Ricci tensor flat,
(iii) Owing to the non-negativity of mass, the
cosmological constant is non-negative definite, and
(iv) Invoking the scalar field paradigm, while the physical
cosmological constant strictly vanishes, it is the dark
energy which resembles a necessarily positive decaying
'cosmological constant'.
And finally, most importantly, we have given a concrete
example how a non-zero cosmological constant can arise
after all in certain normalized gravity theories.
While NGR, at least at the present stage, is not free of its
own delicate points, we hope it will shed some
light on the stubborn cosmological constant puzzle.

\acknowledgments{It is a pleasure to cordially thank our
colleague Ilya Gurwich for valuable and inspiring discussions.
We are grateful to Prof. N. Kaloper for informing us, after the
completion of this work, about the existence of ref.\cite{Tseytlin}.}

\end{document}